\documentclass[preprint,preprintnumbers,nofootinbib,aps,twocolumn,10pt]{revtex4-1}

\usepackage{amsmath,amssymb,bm,epsfig}
\usepackage{color}
\usepackage{natbib}
\usepackage{hyperref} 
\usepackage{ulem}
\usepackage{graphicx}
\RequirePackage{lineno}

\usepackage{xcolor,colortbl}
\usepackage{pifont}

\def \beq{\begin{equation}}
\def \eeq{\end{equation}}
\def \beqa{\begin{eqnarray}}
\def \eeqa{\end{eqnarray}}
\def \la{\langle}
\def \ra{\rangle}
\def \l{\left(}
\def \r{\right)}

\newcommand{\nn}{\nonumber}

\usepackage{xspace}

\newcommand{\sNN}{\sqrt{s_{\rm NN}}}

\begin{document}

\title{Initial condition from the shadowed Glauber model}
 
\author{Sandeep Chatterjee}
\email{sandeepc@vecc.gov.in}
\affiliation{Theoretical Physics Division, 
Variable Energy Cyclotron Centre, 1/AF Bidhannagar, 
Kolkata, 700064, India}

\author{Sushant K. Singh}
\email{sushantsk@vecc.gov.in}
\affiliation{Theoretical Physics Division, 
Variable Energy Cyclotron Centre, 1/AF Bidhannagar, 
Kolkata, 700064, India}

\author{Snigdha Ghosh}
\affiliation{Theoretical Physics Division, 
Variable Energy Cyclotron Centre, 1/AF Bidhannagar, 
Kolkata, 700064, India}

\author{Md Hasanujjaman}
\affiliation{Theoretical Physics Division, 
Variable Energy Cyclotron Centre, 1/AF Bidhannagar, 
Kolkata, 700064, India}

\author{Jane Alam}
\affiliation{Theoretical Physics Division, 
Variable Energy Cyclotron Centre, 1/AF Bidhannagar, 
Kolkata, 700064, India}

\author{Sourav Sarkar}
\affiliation{Theoretical Physics Division, 
Variable Energy Cyclotron Centre, 1/AF Bidhannagar, 
Kolkata, 700064, India}

\begin{abstract}
The two component Monte-Carlo Glauber model predicts a knee-like structure in 
the centrality dependence of elliptic flow $v_2$ in Uranium+Uranium 
collisions at $\sNN=193$ GeV. It also produces a strong anti-correlation 
between $v_2$ and $dN_{ch}/dy$ in the case of top ZDC events. However, none 
of these features have been observed in data. We address these discrepancies 
by including the effect of nucleon shadowing to the two component Monte-Carlo 
Glauber model. Apart from addressing successfully the above issues, we find 
that the nucleon shadow suppresses the event by event fluctuation of various 
quantities, e.g. $\varepsilon_2$ which is in accordance with expectation from 
the dynamical models of initial condition based on gluon saturation physics.
\end{abstract}
\maketitle

One of the major challenges in heavy ion collision (HIC) experiments is to 
comprehend the initial condition (IC). This is an essential requirement to 
extract crucial physical properties of the quark gluon plasma (QGP) phase 
{\it e.g.} the equation of state, the transport coefficients, etc. 
Studies suggest that the largest uncertainties on the extracted value of the 
ratio of the shear viscosity over entropy density arise from the ignorance of 
the IC~\cite{Song:2010mg}. 

Currently, two types of IC models are available. The first is the Monte-Carlo 
Glauber model (MCGM), a geometry based model of the initial distribution of the 
energy deposited in the transverse plane (with respect to the beam 
axis)~\cite{Bialas:1976ed,Kharzeev:2000ph,Miller:2007ri,Broniowski:2007nz}. The 
dynamical input is restricted to the constant nucleon-nucleon cross section 
$\sigma_{NN}$ at given beam energy. This simple model has been fairly successful 
in providing the centrality dependence of various global observables. The second 
type of models which attempt to generate ICs are based on 
QCD~\cite{Eskola:1999fc,Kharzeev:2001yq,Hirano:2005xf,Schenke:2012wb}. A few of 
the recent successful approaches are color glass condensate based IP-Glasma 
model~\cite{Gale:2012rq} and NLO pQCD and gluon saturation based EKRT 
model~\cite{Niemi:2015qia}.

The recent data from STAR on U+U~\cite{Adamczyk:2015obl} and LHC on 
Pb+Pb~\cite{Timmins:2013hq,Aad:2013xma} have rung the death bell for the MCGM. The 
prolate shape of the U nucleus implies a knee-like structure in the centrality 
dependence of elliptic flow $v_2$~\cite{Voloshin:2010ut,Goldschmidt:2015kpa}. 
For the same reason, a strong anti-correlation is expected between $v_2$ and multiplicity 
in the top zero degree calorimeter (ZDC) events~\cite{Schenke:2014tga}. However, 
none of these unique predictions of the two-component MCGM were seen in the data~\cite{Adamczyk:2015obl}. 
Moreover, MCGM predicts a broader event by event (E/E) distribution of flow 
observables $v_n$ than seen in experiments~\cite{Gale:2012rq,Niemi:2015qia}. These 
observations tend to rule out the two component MCGM as a viable candidate 
to provide  IC in HIC. On the other hand, the predictions from the 
dynamical models are in agreement with data~\cite{Gale:2012rq,Niemi:2015qia}. 
There have been several attempts to revive the Glauber model. The absence of knee in 
the centrality dependence of $v_2$ in U+U could be explained by introducing 
fluctuating weight factors for the energy depositing sources~\cite{Rybczynski:2012av}. 
The constituent quark model using wounded constituent quarks as sources for energy 
deposition is in agreement with the U+U data on $v_2$~\cite{Eremin:2003qn, Adler:2013aqf, 
Adamczyk:2015obl}. In Ref.~\cite{Moreland:2014oya}, the simple 
two component scheme was replaced by a reduced nuclear thickness function that yielded 
results in better agreement with data. The simplicity of the two component MCGM 
approach is appealing as well as computationally cheap to implement. Therefore, it 
is an interesting question to ask whether to give up this geometrical idea altogether 
is the only way to make peace with the current data. 

\begin{figure}[]
\includegraphics[scale=0.11]{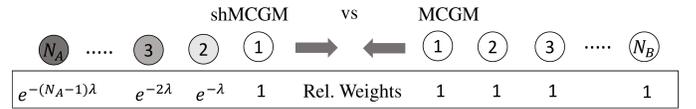}
\caption{The relative weight factors of each participant for energy deposition in the 
shMCGM (left) vs MCGM (right) for the simple case of collision between two rows 
with $N_A$ (right going) and $N_B$ (left going) number of nucleons in each.}
\label{fig.NonN}\end{figure}

In the Eikonal limit the collisions of nucleon rows form the basis of MCGM. 
Fig.~\ref{fig.NonN} illustrates  such a scenario. 
As shown in the right hand side of Fig.~\ref{fig.NonN}, 
we note that in a MCGM approach all the nucleons that contribute to energy 
deposition are treated democratically and hence receive the 
same relative weight for energy deposition. Here we modify this approach by 
not treating all the participants and binary collisions identically -{\it{ the 
contribution to energy deposition by nucleons seated deep inside the nucleus  
is shadowed by those leading in front}}. We call this the shadowed Monte Carlo 
Glauber Model (shMCGM). We use a simple suppression factor $\mathcal{S}\l n,
\lambda\r$ for the contribution from a nucleon which finds $n$ other participants 
from the same nucleus ahead of it 
\beqa
\mathcal{S}\l n,\lambda\r &=& e^{-n\lambda}\label{eq.suppression}
\eeqa
where $\lambda$ is a phenomenological parameter that is to be extracted 
from experiments. The modified weight factors in case of shMCGM in the simple case 
of a row on row collision is illustrated on the left hand side of Fig.~\ref{fig.NonN}. 
This idea of shadowing of nucleons by other nucleons inside a nucleus 
was first discussed about sixty years back~\cite{Glauber:1955qq}. However, barring a 
few studies~\cite{Bzdak:2006qk}, it has been rarely applied in the context of HICs. 
Here, we implement this scheme in nuclear collisions at relativistic energies for 
the first time within a Monte-Carlo approach. Our results indicate that nucleon 
shadowing in HICs has crucial consequences on several observables.

We will now discuss the details of the Glauber model. The nucleus is first 
constructed by generating the nucleons sampled from a Woods-Saxon profile. 
The $i$th nucleon from nucleus A with transverse coordinates $\l x^A_i, 
y^A_i\r$ is made to undergo a binary collision with the $j$th nucleon from 
nucleus B with transverse coordinates $\l x^B_j, y^B_j\r$ when their squared 
transverse separation ${\rho^{AB}_{ij}}$ satisfy the following geometrical criteria
\beqa
{\rho^{AB}_{ij}} &\leq& \frac{\sigma_{NN}}{\pi}
\label{eq.collide}
\eeqa
where ${\rho^{AB}_{ij}}$ is defined as
\beqa
{\rho^{AB}_{ij}} &=& \l x^A_i-x^B_j\r^2 + \l y^A_i-y^B_j\r^2
\label{eq.rAB}
\eeqa
Those nucleons from the colliding nuclei whose transverse coordinates satisfy 
Eq.~\ref{eq.collide} are identified as participants. The sum of the number of 
participants, $N_{part}$  and all the possible binary collisions between them, $N_{coll}$ 
give us the number of energy depositing and particle producing sources in an event in the 
Glauber model. The charged particle multiplicity at mid-rapidity $\left.dN_{ch}/dy\right|_{y=0}$ 
(henceforth we will drop the subscript $y=0$ and refer to it by just $dN_{ch}/dy$) in an 
event is expressed as a sum over all these sources with participant and collision 
sources receiving different weight factors parametrised by the hardness factor $f$ as follows:
\beqa
dN_{ch}/dy &=& \frac{\l1-f\r}{2}\sum_i^{N_{part}}w_i + f\sum_i^{N_{coll}}w_i
\label{eq.MCG2comp_multiplicity}
\eeqa
where $w_i$ is source weight factor. In the simplest case it can be just taken as a constant 
$n_0$ equal to the average $dN_{ch}/dy$ for p-p collision. However, multiplicity distribution 
data in p-p and p-A show that the data is best described by sampling $w_i$s from a negative 
binomial distribution (NBD) $P_{NBD}\l w,n_0,k\r$ with mean $n_0$ and variance 
$\sim\frac{1}{k}$~\cite{Bozek:2013uha}
\beqa
P_{NBD}\l w, n_0,k\r &=& \frac{\Gamma\l k+w\r}{\Gamma\l k\r\Gamma\l w+1\r}\frac{n_0^wk^k}
{\l n_0+k\r^{w+k}}
\label{eq.pnbd}\eeqa
In this work we follow the same strategy and consider NBD fluctuations in the $w_i$s to generate 
multiplicity. The energy deposition ansatz is also similar to Eq.~\ref{eq.MCG2comp_multiplicity}. 
However, for the energy deposition scheme we have not considered any additional fluctuation as in 
Ref.~\cite{Bozek:2013uha} and just considered $w_i=\epsilon_0$, a constant. We will now turn our 
discussion towards shMCGM. The only modification that is made to the above framework to obtain 
shMCGM is in Eq.~\ref{eq.MCG2comp_multiplicity} where there are now additional weight factors 
$\mathcal{S}\l n_i,\lambda\r$ defined in Eq.~\ref{eq.suppression} due to shadowing. $n_i$ depends on 
whether the source is a participant or a binary collision: in case of a participant source it is equal 
to the number of nucleons that shadow it and in case of a binary collision it is the sum of the 
number of nucleons that shadow the two participants of the given binary collision. 

We will now first try to understand the effect of introducing shadowing 
in shMCGM on observables computed thereof qualitatively neglecting the NBD fluctuations 
of the sources before discussing the Monte-Carlo results. Let us consider two nuclei 
$A$ and $B$ approaching each other along the z axis with the origin at the center of mass 
of the two nuclei system. As discussed earlier, the contribution to energy deposition 
due to a participant nucleon $i$ from the nucleus $A$ at $\l x^i_A, y^i_A, z^i_A \r$ in 
the two-component scheme occurs through the participant term ${N_i}_{part}^{sh}$ and 
collision term ${N_i}_{coll}^{sh}$ which are obtained as follows
\beqa
n\l x^i_A, y^i_A, z^i_A\r &=&
\sum_{j=1,j\neq i}^{N_{A}}\Theta\l\frac{\sigma_{NN}}{\pi}-{\rho_{ij}^{AA}}\r\nn\\
&&\times\Theta\l|z^i_A| - |z^j_A|\r\label{eq.n}\\
m\l x^i_A, y^i_A, z^i_A\r &=& 
\sum_{j=1}^{N_B}\Theta\l\frac{\sigma_{NN}}{\pi}-{\rho_{ij}^{AB}}\r\label{eq.m}\\
{N_i}_{part}^{sh}\l x^i_A, y^i_A, z^i_A, \lambda \r &=& \mathcal{S}\l n,\lambda\r\label{eq.npartsh}\\
{N_i}_{coll}^{sh}\l x^i_A, y^i_A, z^i_A, \lambda \r &=& \mathcal{S}\l n,\lambda\r\sum_{j=0}^{m-1}
\mathcal{S}\l j,\lambda\r\label{eq.ncollsh}
\eeqa
Here $\Theta\l x\r=1$ for $x\geq0$ and $0$ for $x<0$. $n$ has been already introduced 
in Eq.~\ref{eq.suppression} and $m$ is the number of nucleons from nucleus B 
with which the $i$th nucleon from nucleus A collides.

In order to understand the nucleon shadowing effect we will now focus on our 
earlier simple case illustrated in Fig.~\ref{fig.NonN}. Here
\beqa
N_{part} &=& N_A + N_B\label{eq.npartex}\\
N_{coll} &=& N_AN_B\label{eq.ncollex}\\
N_{part}^{sh} &=& \sum_{i=0}^{N_A-1} \mathcal{S}\l i,\lambda\r + 
\sum_{i=0}^{N_B-1} \mathcal{S}\l i,\lambda\r\nn\\
&=& \frac{2-e^{-\lambda N_A}-e^{-\lambda N_B}}{1-e^{-\lambda}}\label{eq.npartshex}\\
N_{coll}^{sh} &=& \frac{1-e^{-\lambda N_A}-e^{-\lambda N_B}+e^{-\lambda\l N_A+N_B\r}}
{\l 1-e^{-\lambda} \r^2}\label{eq.ncollshex}
\eeqa
The source for the E/E fluctuations in the MCGM ICs lie in the E/E 
fluctuations of the positions of the nucleons sampled from the Woods-Saxon 
profile of the nucleus which in this case translate into fluctuation in 
$N_A\l x,y\r$ and $N_B\l x,y\r$
\beqa
\delta N_{part} &=& \delta N_A + \delta N_B\label{eq.delnpartex} \\
\delta N_{coll} &=& N_B\delta N_A + N_A\delta N_B\label{eq.delncollex} \\
\delta N_{part}^{sh} &=& \frac{\lambda}{1-e^{-\lambda}}\l e^{-\lambda N_A}\delta N_A + 
 e^{-\lambda N_B}\delta N_B \r\label{eq.delnpartshex}\\
\delta N_{coll}^{sh} &=& \frac{\lambda}{\l 1-e^{-\lambda}\r^2}\l\l 1-e^{-\lambda N_B}\r 
e^{-\lambda N_A}\delta N_A\right.\nn\\
&&\left.+ \l 1-e^{-\lambda N_A}\r e^{-\lambda N_B}\delta N_B\r\label{eq.delncollshex}
\eeqa
As seen from Eqs.~\ref{eq.npartshex}, \ref{eq.ncollshex}, \ref{eq.delnpartshex} 
and \ref{eq.delncollshex} in the $\lambda\rightarrow\infty$ limit, 
$N_{part}^{sh}\rightarrow2$, $N_{coll}^{sh}\rightarrow1$ while their fluctuation is 
completely suppressed, thus in this limit
the energy is deposited by the leading nucleons alone,  
irrespective of the number of nucleons present in the colliding rows. 
However, as it turns out from fits to data, the more relevant case is the 
$\lambda\rightarrow0$ limit. With $N_A\sim N_B\sim N$, 
$\delta N_A\sim \delta N_B\sim \delta N$ and $\l \lambda, \lambda N\r<<1$, we get
\beqa
N^{sh}_{part} &\simeq& \l1 -\frac{\l N-1\r}{2}\lambda\r N_{part}\label{eq.npartshsmalll}\\
N^{sh}_{coll} &\simeq& \l1 -\l N-1\r\lambda\r N_{coll}\label{eq.ncollshsmalll}\\
\delta N_X^{sh} &\simeq& \l 1-N\lambda\r\delta N_X\label{eq.nflucshsmalll}
\eeqa
where $N_X=\l N_{part}, N_{coll}\r$. Thus the shadow effect is expected to suppress 
the energy deposited as well as the E/E fluctuations of observables.

\begin{table}[b]
\begin{center}
\begin{tabular}{|c|c|c|c|c|c|}
\hline
Model & system & $n_0$ & $k$ & $f$ & $\lambda$ \\
\hline
MCGM & Au+Au & 2.37 & 1.1 & 0.14 & -\\
\hline
MCGM & U+U & 2.30 & 1.1 & 0.14 & -\\
\hline
shMCGM & Au+Au & 2.83 & 1.1 & 0.32 & 0.12\\
\hline
shMCGM & U+U & 2.83 & 1.1 & 0.32 & 0.12\\
\hline
\end{tabular}
\end{center}
\caption{The values of the parameters of the Glauber model used in this work. 
The results from a detailed fit procedure with allowed range in the parameter 
space will be reported elsewhere.}
\label{tab.fit}\end{table}

\begin{figure}[]
\begin{center}
\includegraphics[angle=-90, scale=0.28]{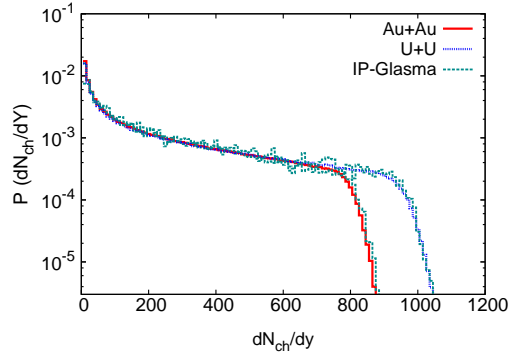}
\end{center}
\caption{The multiplicity distribution plot for U+U at $\sNN=193$ GeV and 
Au+Au at $\sNN=200$ GeV in shMCGM. The IP-Glasma data is from Ref.~\cite{Schenke:2014tga}.}
\label{fig.pdnchdy}\end{figure}
We will now report our findings. In this work we use the same Woods-Saxon parameters 
for U and deformed Au as used in Ref.~\cite{Schenke:2014tga}. The parameters of the 
MCGM as well as shMCGM are fixed by comparing the variation of the probability distribution 
for multiplicity, $P\l dN_{ch}/dy\r$ with $dN_{ch}/dy$ to that of IP-Glasma~\cite{Schenke:2014tga} 
as shown in Fig.~\ref{fig.pdnchdy}. The values of the parameters used in this work are 
mentioned in Table~\ref{tab.fit}. From MCGM to shMCGM, the nucleon shadow effect in energy 
deposition is compensated by increasing the value of $n_0$ and $f$. The fits to multiplicity 
distribution in p-Pb collisions at $\sNN=5.02$ TeV yield $k\sim1$~\cite{Bozek:2013uha}. In this 
work, we have taken $k=1.1$.

We also compute the eccentricities of the initial energy density deposited that are expected 
to drive the momentum anisotropies of the produced particles measured in experiments. The 
standard way to compute the eccentricities $\varepsilon_n$ of the overlap region is
\beqa
\varepsilon_ne^{i\Psi_n} &=& \frac{\la r^ne^{in\phi_n}\ra}{\la r^n\ra}
\label{eq.ecc}
\eeqa
where the averages $\la ... \ra$ are taken with the initial energy distribution on the 
transverse plane $\epsilon\l x,y\r$ as the weight function. The energy $\epsilon_i\l x,y\r$ 
deposited by the $i$th source of strength $\epsilon_0$ located at $\l x_i,y_i\r$ is 
smeared by a Gaussian profile  
\beqa
\epsilon_i\l x,y\r &=& \frac{\epsilon_0}{2\pi\sigma^2}e^{-\frac{\l x-x_i\r^2+\l y-y_i\r^2}
{2\sigma^2}}
\label{eq.energy_single_source}
\eeqa
where we have set the width $\sigma=0.6$ $\text{fm}$.

The full overlap U+U collision configurations can be of two types: (i) Body-Body (BB)- 
in this case the U nuclei approach each other along their minor axes and the overlap 
region has higher ellipticity, and (ii) Tip-Tip (TT)- in this case the U nuclei 
approach along their major axes and the overlap region is circular and has smaller 
$\varepsilon_2$ as compared to the BB configuration~\cite{Haque:2011aa}. While both 
the configurations have similar $N_{part}$, $N_{coll}$ is larger by a factor 
$\sim1.3$ in the case of TT events as compared to BB. Thus the two component scheme 
as given by Eq.~\ref{eq.MCG2comp_multiplicity} predicts that the highest multiplicity 
U+U events must be from TT configurations with smaller $v_2$ due to circular shape
of the collision zone. As discussed 
earlier this leads to the prediction of a knee-like structure in the  centrality 
variation of $v_2$ as well as a strong anti-correlation of $v_2$ vs multiplicity 
in the top ZDC events, unlike what is seen in experiments.

\begin{figure}[]
\includegraphics[angle = -90, scale = 0.28]{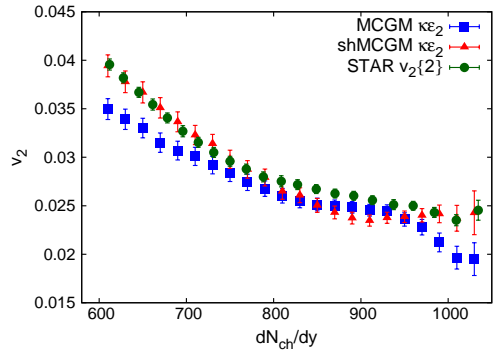}
\caption{The centrality dependence of $v_2$ for U+U at $\sNN=193$ GeV as obtained in 
MCGM and shMCGM. The experimental data is from Ref.~\cite{Adamczyk:2015obl}.}
\label{fig.e2}\end{figure}
The $\varepsilon_2$ calculated within the ambit of MCGM and shMCGM have been used to 
obtain $v_2$ through the scaling $v_2=\kappa\varepsilon_2$ with $\kappa\sim0.2$~\cite{Kolb:2000sd}.
In Fig.\ref{fig.e2}, we have compared the centrality dependence of this scaled 
$\varepsilon_2$ with the STAR data on $v_2\{2\}$ in U+U collisions~\cite{Adamczyk:2015obl}. 
We find a fairly good qualitative agreement between data and shMCGM. The 
most striking difference between shMCGM and MCGM occurs for the most central events 
where the effect of shadow is expected to be the highest. Clearly, the knee-like 
structure that is predicted by the MCGM (but not observed in data) is washed away 
in the case of shMCGM which accurately follows the data. The knee like structure has 
vanished because the effect of shadowing moderates the collision process to bring a 
balance  in the effective numbers of collisions for tip-tip and body-body geometry by 
reducing $N_{coll}$ more in TT compared to BB configurations.

Another interesting point to note from Fig.~\ref{fig.e2} is that $\varepsilon_2$ in case of 
shMCGM is higher as compared to MCGM. In a typical mid-central collision where $\varepsilon_2$ 
is generated mainly because of the elliptical shape of the overlapped region, the ends of 
the major (minor) axis of the elliptical overlap region receive contribution from the 
boundary region of both (one) nucleus. Hence there is lesser (higher) energy deposition. 
Now as seen in Eqs.~\ref{eq.npartshsmalll} and \ref{eq.ncollshsmalll}, the effect of 
shadow is weaker (stronger) where lesser (more) nucleons are expected to deposit energy. 
This leads to milder (stronger) shadowing effect at the ends of the major (minor) axis, 
which effectively enhances the ellipticity in shMCGM compared to MCGM. Similar arguments 
also show that models based on gluon saturation physics are expected to generate higher 
$\varepsilon_2$ as compared to MCGM~\cite{Drescher:2006pi}.

\begin{figure}[]
\begin{center}
\includegraphics[angle = -90, scale = 0.27]{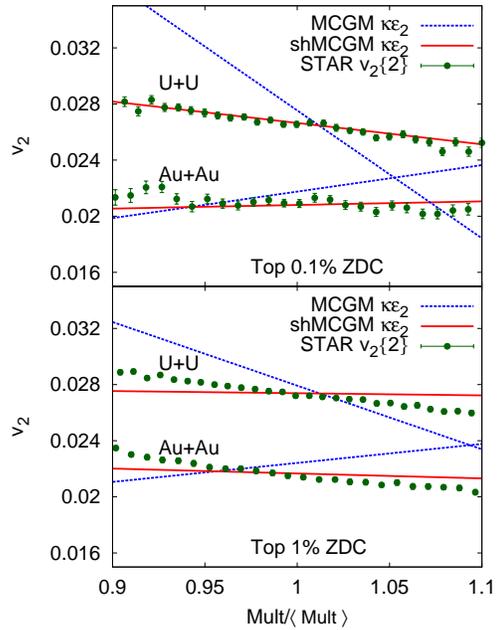}
\end{center}
\caption{Scaled $\varepsilon_2$ vs multiplicity plot for U+U at $\sNN=193$ GeV and 
Au+Au at $\sNN=200$ GeV for top ZDC events as obtained in MCGM and shMCGM. The 
experimental data is from Ref.~\cite{Adamczyk:2015obl}.}
\label{fig.zdc}\end{figure}
The high multiplicity events have been looked at in yet another way- by applying 
cuts on the ZDC. Top $\l 0-0.1\r\%$ and $\l 0-1\r\%$ ZDC events have been analyzed 
in STAR~\cite{Adamczyk:2015obl}. The MCGM predicts a strong anti-correlation between 
$v_2\l\kappa\varepsilon_2\r$ and multiplicity in these events. However, such strong 
anti-correlation was not seen in experiments as well as in the IP-Glasma model 
~\cite{Schenke:2014tga}. Here, we find that the new phenomenological 
parameter $\lambda$ which encapsulates the nucleon shadowing effect could be suitably 
tuned in the shMCGM to agree with data. For the same value of $\lambda$, a much 
improved agreement is seen between data and shMCGM in Fig.~\ref{fig.zdc} for both 
collision systems, U+U and Au+Au.

\begin{figure}[]
\begin{center}
\includegraphics[angle = -90, scale = 0.27]{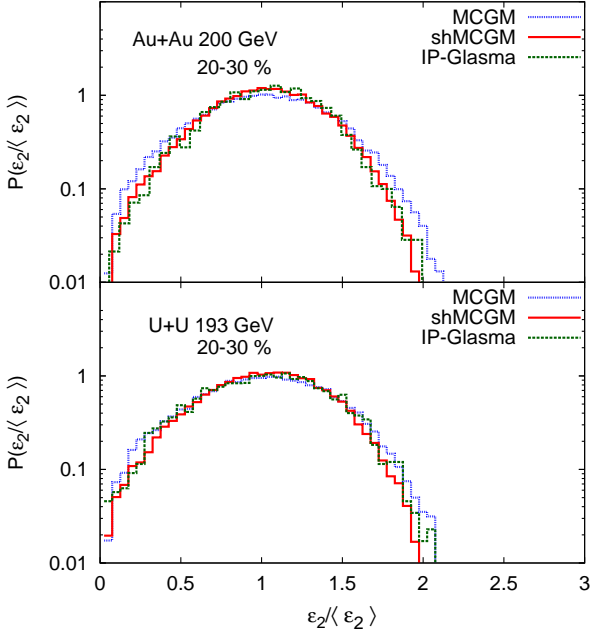}
\end{center}
\caption{The event by event probability distribution of $\varepsilon_2$ for U+U at 
$\sNN=193$ GeV and Au+Au at $\sNN=200$ GeV for $\l20-30\r\%$ centrality events as 
obtained in MCGM and shMCGM. The IP-Glasma data is from Ref.~\cite{Schenke:2014tga}.}
\label{fig.ebye}\end{figure}
We finally turn our attention towards E/E distribution of $\varepsilon_2$. It has 
already been pointed out that the E/E distribution of $\varepsilon_2$ in dynamical 
models is narrower than that of MCGM~\cite{Gale:2012rq,Schenke:2014tga,Niemi:2015qia}. 
We noted earlier in Eq.~\ref{eq.nflucshsmalll} that the shadow suppresses both 
$N_{part}$ and $N_{coll}$ as well as their fluctuations. This suggests that E/E 
fluctuation of other observables computed in the shMCGM should be narrower compared 
to the MCGM. In Fig.~\ref{fig.ebye} we have plotted the E/E distribution of 
$\varepsilon_2/\la\varepsilon_2\ra$ for U+U and Au+Au collisions for $\l20-30\r\%$ 
centrality. The E/E distribution from the shMCGM is indeed narrower compared 
to the MCGM case and compares well with the IP-Glasma distribution. Thus, overall we 
find the shMCGM to provide ICs for U+U at $\sNN=193$ GeV and Au+Au at $\sNN=200$ GeV 
that are in agreement with experimental data as well as with predictions from other 
dynamical models. It is interesting to note that the results of shMCGM are in good 
agreement with data for Pb+Pb collisions at $\sNN=2.76$ TeV ~\cite{Ghosh:2016npo}.

Finally to conclude, MCGM has been employed routinely in the study of HICs to ascertain 
ICs, based on geometric considerations and a few free parameters that are extracted from 
fits to data. However, lately it has been superseded by the QCD based models of 
ICs. The high multiplicity U+U events and the data on E/E flow distributions at the 
LHC have clearly ruled out the simple geometrical MCGM in favour of the latter models. 
Here we explored the possibility of bridging the above gap by including the effect of 
shadowing due to leading nucleons in the MCGM, an idea that dates back to 
$1950$s~\cite{Glauber:1955qq} but rarely explored in HICs. In the conventional MCGM, all the 
sources are given equal weightage to deposit energy. In the shMCGM, leading sources 
are given a larger weightage than those in the interior. For a suitable choice of the 
shadow parameter, we find good agreement between data and shMCGM. We also argue that 
the presence of shadow in the shMCGM invariably reduces the E/E fluctuation as compared 
to the conventional MCGM. Thus, we now find good agreement between IP-Glasma and shMCGM 
predictions of the E/E distribution of $\varepsilon_2$. Our study hints that incorporating 
shadowing in the MCGM could bring such geometrical models closer to dynamical models. At 
this point it is an interesting question to ask whether our ansatz for shadow given by 
Eq.~\ref{eq.suppression} is unique or equally good description of the same observables is 
possible with other ansatz. In this regard it might be a worthy exercise to check whether 
on starting from a QCD based dynamical approach it is possible to derive an effective form 
of the shadowing factor and predict the shadow parameter $\lambda$ in terms of the relevant 
scales of the problem rather than treating it as a free parameter. 

{{\it Acknowledgement:}} We would like to thank Prithwish Tribedy for providing the IP-Glasma 
data. SC acknowledges him for many fruitful discussions on the initial condition and thanks 
``Centre for Nuclear Theory" [PIC XII-R$\&$D-VEC-5.02.0500], Variable Energy Cyclotron Centre 
for support. SG and MH acknowledge Department of Atomic Energy, Govt. of India for support.
\bibliographystyle{apsrev4-1} 
\bibliography{Shadow}

\end{document}